\documentstyle[12pt]{article}

\begin{document}
\begin{center}
\Large
{\bf CCD Photometric Observations of the Galactic Globular Clusters NGC 1904 and NGC 6341}

\vspace*{0.5in}
\large
{\bf Alok C. Gupta}
\end{center}

\noindent
\normalsize
Harish-Chandra Research Institute, Chhatnag Road, Jhunsi, Allahabad $-$ 211 019,
India  \\
{\bf email:} agupta@mri.ernet.in

\vspace*{0.2in}
\noindent
\Large
{\bf ABSTRACT} \\
\normalsize
We present multi-colour broad band CCD photometric observations of a few thousands
of stars located region of the galactic globular clusters NGC 1904 and NGC 6341. 
These observations are used to generate their deep colour magnitude diagrams (CMDs) to
 study  overall morphologies of sub giant, red giant, asymptotic giant and 
horizontal branches. We also determine the main sequence morphology of the NGC 
6341. A comparison of their CMDs with the theoretical isochrones given by 
Demarque et al. (1996) indicates an age of 14 and 16 Gyr
for NGC 1904 and NGC 6341 GGCs respectively. \\ 
We present a new method for artificial add star experiment. Using this method on the 
photometric data of galactic globular cluster NGC 6341, we determined the 
completeness factor in B, V and I pass bands. We have determined the luminosity function 
of the main sequence stars down to V $\sim$ 21.0 (M$_{V} \sim$ 6.5) from the V Vs 
(B $-$ V) colour magnitude diagram of this cluster.  \\
Using accurate parameters of 64 GGCs, we derived 
relations between galactocentric distance R, distance from the galactic plane 
Z$_{g}$ and metalicity [Fe/H] of GGCs. \\

\noindent
{\it PACS:} 

\newpage
\section {INTRODUCTION} 
Precise multi-colour photometric observations of the Galactic Globular Clusters (GGCs)
provide valuable information about the late stages of stellar evolution in low-mass
stars, as they are amongst the oldest objects in our galaxy. They are also useful for
the study of galactic evolution. We have selected two GGCs namely NGC 1904 (M 79), and 
NGC 6341 (M 92) for the present study and provide basic informations about them in Table 1.
A brief description of their earlier photometric studies is given below.

\noindent
{\bf Table 1.} Basic cluster parameters of the four galactic globular clusters
included in the present study. The GGCs parameters are taken from Djorgovski (1993).  
\begin{center}
\begin{tabular}{lcc} \\\hline
{\bf Parameter} &   {\bf Cluster} &     \\\cline{2-3}
                & NGC 1904 & NGC 6341 \\\hline
R. A. (2000.0)  & 05$^{h}$ 24$^{m}$ 10.6$^{s}$ & 17$^{h}$ 17$^{m}$ 07.3$^{s}$ \\
Dec. (2000.0)   & $-24^{0} 31^{'} 27^{"}$ & $+43^{0} 08^{'} 11^{"}$ \\
Gal. longitude (deg) & 227.229 & 68.339 \\
Gal. latitude (deg)  & $-$29.351 & 34.859 \\
E(B$-$V) (mag) & 0.01 & 0.02 \\
Distance (kpc) & 13.0 & 7.5 \\
Concentration c & 1.72 & 1.81 \\
log (R) (kpc) & 1.28 & 0.96 \\
log (Z$_g$) (kpc) & 0.80 & 0.63 \\
{[Fe/H]} & $-$1.69 & $-$2.24 \\\hline
\end{tabular}
\end{center}

\subsection {NGC 1904}

First photoelectric photometric study of the cluster NGC 1904 is done by Goranskij (1976)
and in the same year Alcaino (1976) published photographic BV colour-magnitude diagram 
(CMD) of 95 stars. The main features of the CMD are: the presence of a large sample of blue HB stars in contrast to
lack of red stars. The values of reddening E(B$-$V) and true distance modulus (m$-$M)$_{0}$
are determined as 0.09 mag and 15.28 mag  respectively. Stetson \& Harris (1977) published UBV
photographic data of 152 stars in the magnitude range 13.0 $<$ V $<$ 17.2. They found
very small value of foreground reddening E(B$-$V) = 0.01 $\pm$ 0.01 mag and reported 
the location of the HB at V$_{HB} =$ 16.2 $\pm$ 0.1 mag. The cluster is therefore located 
at a geocentric distance of 13 kpc and at a galactocentric distance of 21 kpc. \\
First main-sequence (MS) photometry of NGC 1904 is presented by Harris et al. (1983) 
using SIT Vidicon Camera. They produced a BV CMD for 264 stars in the magnitude range 16 $<$ V $<$ 22. They found
the MS turn off at V $\sim$ 19.4 mag and using theoretical isochrones, derived age of the
cluster between 12 to 18 Gyr. Cordoni \& Auriere (1983) presented BV magnitudes for 161
stars in a 1$^{'} \times$ 1$^{'}$ field centered on the cluster.   \\
First CCD photometry of this cluster is carried out by Heasley et al. (1986). Their BV CCD
photometric study upto V $\sim$ 21.5 mag is well below the MS turn off (MSTO). They
found that the cluster is located at 20 kpc away from the galactic center. They derived
cluster age as 16 Gyr. Gratton \& Ortolani (1986) carried out BV CCD photometric study of
227 stars upto V $\sim$ 21.5 mag and derived the cluster age as 15 to 18 Gyr. \\
Ferraro et al. (1992) presented BV photographic photometry of 3188 stars down to V $\sim$ 21
mag. They studied the overall morphology of different branches of CMD of the cluster. They
located RGB bump at V $=$ 16.00 $\pm$ 0.05 mag and determined the mean metalicity, [Fe/H]
$= -$1.60 $\pm$ 0.20; Horizontal branch (HB) at V$_{HB} =$ 16.15 $\pm$ 0.10 mag; MSTO
at V$_{TO} =$ 19.60 $\pm$ 0.20 mag; $\Delta V_{TO-HB} =$ 3.45 $\pm$ 0.22 mag; M$_{V}$(HB)
= 0.70 $\pm$ 0.07 mag, (m-M)$_{V} =$ 15.45 $\pm$ 0.02 mag which put the cluster at a
distance of 12$\pm$2 kpc from the Sun. \\
BVRI CCD photometry of this cluster is done by Alcaino et al. (1994). They determined
V$_{TO} =$ 19.60 $\pm$ 0.10 mag with turn off colours at B$-$V = 0.40, V$-$R = 0.27 and
V$-$I = 0.57 all with estimated external errors of $\pm$0.06 mag. By fitting the isochrones,
they estimated the cluster age as 16 Gyr. Recently UBV CCD photometric study of the cluster
NGC 1904 is done by Kravtsov et al. (1997). They determined MS turn off at V = 19.70 $\pm$
0.05 mag, B$-$V = 0.415 $\pm$ 0.01 mag. The horizontal branch level at the blue edge of the
instability strip has V = 16.25 $\pm$ 0.10 mag. They determined the cluster metalicity
[Fe/H] $= -$ 1.76 $\pm$ 0.20 and age 15 $-$ 19 Gyr. \\
The cluster parameters derived in earlier studies are given Table 2.

\noindent
{\bf Table 2.} The cluster NGC 1904 parameters determined in previous photometric studies
as follows:

\scriptsize
\begin{tabular}{lllllcl} \\\hline
(m$-$M)$_{V}$ & E(B$-$V) & V$_{TO}$ & (B$-$V)$_{TO}$ & V$_{HB}$ & Age (Gyr) & References  \\ \hline
15.55  & 0.09 &   &  & & & Alcaino et al. (1976) \\
       & 0.01$\pm$0.01 & & & 16.2$\pm$0.1 &  & Stetson \& Harris (1977) \\
15.60 & 0.02 & 19.4 & 0.52 & 16.2 & 12$-$18 & Harris et al. (1983) \\
15.65 & 0.01 &   &  &  & 16 & Heasley et al. (1986) \\
      &    & 19.6$\pm$0.1 & 0.49 &  & 15$-$18 & Gratton \& Ortolani (1986) \\
15.45$\pm$0.02 & 0.01 & 19.6$\pm$0.2 & & 16.15$\pm$0.10 & & Ferraro et al. (1992) \\
  &  & 19.6$\pm$0.1 & 0.40$\pm$0.06  &  & 16 & Alcaino et al. (1994) \\
  &  & 19.70$\pm$0.05 & 0.415$\pm$0.01 & 16.25$\pm$0.10 & 15$-$19 & Kravtsov et al. (1977) \\
15.63 & 0.01 &  &  & 16.27$\pm$0.07 &  & Ferraro et al. (1999) \\\hline
\end{tabular}
\normalsize

\subsection {NGC 6341}

Being one of the most metal poor globular clusters in the Galaxy ([Fe/H] $= -$2.24;
Djorgovski 1993), it provides very good opportunity for studying old and extremely
metal poor stars. \\
Previous photometric studies are those by Arp, Baum \& Sandage (1952, 1953), Sandage
\& Walker (1966), Sandage (1970), Sandage \& Katem (1983), Sandage (1983). CCD based
photometric studies of this cluster is done by Heasley \& Christian (1986) and 
Stetson \& Harris (1988). 
The cluster parameters derived by them are given in Table 3. VI CCD photometric study
has been done recently by Johnson \& Bolte (1998) for this cluster. They reported the
values for turn-off magnitude V$_{TO} =$ 18.80, colour of turn-off point (V$-$I)$_{TO} =$ 0.56
and the magnitude of blue edge of instability strip V$_{HB} =$ 15.20. \\
The reddening to the cluster determined by various authors agrees very well with each
other. However, the distance and age estimates differ significantly. \\
GGCs luminosity function has been normally presented by a Gaussian
distribution. The mean absolute magnitude M$_{V} \sim -$7.3 magnitude
and dispersion $\sigma (M_{V}) \sim$ 1.1 magnitude (Harris and Racine
1979; Hanes and Whittaker 1987; Racine and Harris 1992; Secker and
Harris 1993). However, van den Bergh (1985) has reported that the
observed luminosity function of GGCs is slightly asymmetric with a
long tail extending upto the faint magnitudes. Recently, Secker (1992)
and Racine and Harris (1992) have suggested that GGCs luminosity function
is non-Gaussian and attempted to model it as a t-distribution. GGCs
luminosity function expressed in terms of clusters per unit luminosity
as a truncated power-law model. It is derived by assuming a mass function
constructed from three power laws (McLaughin 1994; Harris and Pudritz
1994). \\
{\bf Table 3.} The parameters of NGC 6341 determined earlier.

\begin{center}
\begin{tabular}{clccl} \\\hline
(m$-$M)$_{V}$ & E(B$-$V) & V$_{TO}$ & (B$-$V)$_{TO}$ &  References  \\ \hline
14.62 & 0.00 & & & Sandage \& Walker (1966) \\
14.63 & 0.02 &  18.43 & & Sandage (1970) \\
14.42 & 0.025$\pm$0.01 & & & Sandage \& Katem (1983) \\
14.42 & 0.02 & & & Sandage (1983) \\
14.60 & 0.02 & & & Heasley \& Christian (1986) \\
14.60 & 0.02 & 18.77 & 0.394 & Stetson \& Harris (1988) \\\hline
\end{tabular}
\end{center}

\vspace*{0.3in}
\noindent
This paper is structured as follows: section 2 describes photometric observations and data
reductions, section 3 about the colour magnitude diagrams of GGCs NGC 1904 and NGC 6341, 
section 4 gives the detail information about our artificial add star experiment, section 5
reports luminosity function of GGC NGC 6341, section 6 presents the ages of these GGCs, 
section 7 describes the relation between GGCs metalicity and their structural parameters
and section 8 reports the conclusions of the present work.

\section {OBSERVATIONS AND DATA REDUCTIONS}

The observations of NGC 1904 were carried out in B, V Johnson and R, I Cousins photometric
passbands using an TK 1024AB2 CCD detector at the f/3.23 prime focus of 2.34 meter Vainu
Bappu Telescope at Vainu Bappu Observatory, Kavalur, India. Each pixel of 1024 $\times$
1024 size CCD is a square of 24 $\mu$m size and the entire chip covers a field
of $\sim$ 11 $\times$ 11 arcmin$^{2}$ on the sky. Bias frames are taken at a regular interval.
Twilight sky flats were taken for correcting the variation in pixel to pixel response. 
The read out
noise for the CCD system was $\sim$ 16 electrons pixel$^{-1}$, while electrons 
per ADU was $\sim$
4. The details of cluster observations are given in Table 4. \\ 
The observations of NGC 6341 were carried out in B, V Johnson and I Cousins
photometric pass bands using RCA SID 501 thinned back illuminated CCD detector
at f/3.29 prime focus of the 2.3 meter Issac Newton Telescope (INT) at
La Palma, Canary Islands, Spain. The cluster region was imaged on the nights 
of 1988 July 21, 22 and
23. In order to avoid saturation of the CCD due to bright stars, we offset
the imaged cluster region. It is $\sim$ 4.5 arcmin south from the cluster
center. Bias, dark, flat fields, photoelectric standards and other programme
clusters regions were also taken on these nights. Nights were good photometric
quality. During the observing run of NGC 6341, seeing was 
$\sim$ 1.1 to 1.2 arcsec. At the prime focus, a pixel of 320 $\times$ 512 size
CCD corresponds to 0.74 arcsec square and the entire chip covers a field of 
view $\sim$ 4.0$\times$6.3 arcmin$^{2}$ on the sky. The readout noise for the 
system $\sim$ 60 electrons pixel$^{-1}$, while electrons per ADU was $\sim$ 4.\\
Flat field exposures ranging from 1 to 10 seconds in each filter were made of
twilight sky. Nine Landolt (1983) photoelectric standards covering a range in
brightness ($10 < V < 12.75$) as well as in colour ($-0.19 < (V-I) < 1.41$)
were observed for calibration purpose. The detail of cluster observation is
given in table 4. Further details of the instrument and observing procedures
have been given in Sagar \& Griffiths (1991, 1998a, 1998b) and Gupta et al. 
(2000). \\
Initial processing of the data frames of NGC 1904 and NGC 6341 were done in the usual 
manner using standard routines
in IRAF package. The evenness of the flat field frames (summed for each colour band) is better
than a few percent in all the filters. The magnitude estimation on each frame has been done
using DAOPHOT II and ALLSTAR II profile fitting software (Stetson 1987, 1992). The stellar
point spread function (PSF) was evaluated using Penny model of DAOPHOT II from several
uncontaminated stars present in each frame. The image parameters and errors provided by
DAOPHOT II and ALLSTAR II were used to reject poor measurements. About 10 \% stars were
rejected in this process. After all the frames were reduced, Stetson's (1992) DAOMATCH
and DAOMASTER routines were used for cross identifying the stars measured on different
frames of the same cluster region. 

\noindent
{\bf Table 4.} Log of Observations of NGC 1904 and NGC 6341

\begin{tabular}{cccl} \\\hline
Cluster & Date   & Filter & {No. of Exposures $\times$ Exposure Time (in sec.)} \\\hline
NGC 1904 & 1996 Jan. 13 & B  & 1 $\times$ 120, 1 $\times$ 300, 2 $\times$ 1200 \\
         & 1996 Jan. 13 & V  & 1 $\times$ 60, 1 $\times$ 180, 2 $\times$ 900 \\
         & 1996 Jan. 13 & R  & 1 $\times$ 30, 1 $\times$ 60, 2 $\times$ 600 \\
         & 1996 Jan. 13 & I  & 1 $\times$ 30, 1 $\times$ 60, 1 $\times$ 420, 1 $\times$ 600 \\
NGC 6341 & 1988 July 21 & B  & 1 $\times$ 20, 3 $\times$ 30, 2 $\times$ 110, 1 $\times$ 220 \\
         & 1988 July 21 & V  & 1 $\times$ 10, 4 $\times$ 20, 1 $\times$ 40, 2 $\times$ 100, 1 $\times$ 200 \\
         & 1988 July 21 & I  & 4 $\times$ 15, 3 $\times$ 150, 1 $\times$ 250 \\
         & 1988 July 22 & B  & 2 $\times$ 20, 4 $\times$ 120 \\
         & 1988 July 22 & V  & 2 $\times$ 20, 4 $\times$ 100 \\
         & 1988 July 22 & I  & 2 $\times$ 15, 4 $\times$ 150 \\
         & 1988 July 23 & B  & 1 $\times$ 20, 1 $\times$ 220 \\
         & 1988 July 23 & V  & 1 $\times$ 20, 1 $\times$ 220 \\
         & 1988 July 23 & I  & 1 $\times$ 15, 1 $\times$ 150 \\\hline
\end{tabular}

\vspace*{0.3in}
\noindent
The CCD instrumental magnitude of stars in NGC 1904 have been calibrated using local 
standards in NGC 1904 field (Alcaino et al. 1987). The colour equations for this GGC 
are given below.
In the following equations b, v, r, i represent instrumental magnitudes while B, V, R, I
represent standard magnitudes in B, V, R, I passbands respectively.
\begin{eqnarray}
B = b + (-0.2745\pm0.019) (b-v)  + (3.236\pm0.028) \\
V = v + (-0.0102\pm0.006) (v-i)  + (2.794\pm0.008) \\
R = r + (-0.0406\pm0.038) (v-r)  + (2.622\pm0.024) \\
I = i + (-0.0531\pm0.040) (v-i)  + (2.677\pm0.051)
\end{eqnarray}

\noindent
The CCD instrumental magnitudes of stars in NGC 6341 have been 
calibrated using the colour equations given by Sagar \& Griffiths (1991), as 
the present observations were taken with same equipment during the same period. 
During observations, the values of atmospheric extinction coefficients
in the V passband is determined by Carlsberg Automatic meridian circle
between 0.10 and 0.11 mag per unit airmass with almost negligible
($\sim -$0.003 mag) hourly rate of change of extinction. These along
with mean (B$-$V) atmospheric extinction coefficients for the site were
used in determining the colour equations for the CCD system using Landolt
(1983) standards (Sagar \& Griffiths 1991). 
Zero points for the B, V and I pass bands cluster frames were determined
with respect to nine photoelectric standards of Landolt (1983) by taking
into the account the differences in exposure time, atmospheric extinction
coefficients and difference between aperture and PSF magnitudes. The zero
points are uncertain by 0.02 magnitude in B, V and I pass bands. Errors 
in the present photometric observations of NGC 1904 and NGC 6341 as a function
of brightness are plotted in figure 1. 

\section {COLOUR MAGNITUDE DIAGRAMS} 

\subsection {NGC 1904}

\subsubsection {Overall Morphology of NGC 1904}

The V, (B$-$V); V, (V$-$R) and V, (V$-$I) apparent CMDs of 4402 stars observed by us
with DAOPHOT II photometric errors in V, (B$-$V), (V$-$R) and (V$-$I) less than 0.07
mag are presented in figure 2. Our CMDs gives the value of turn off mag V$_{TO} =$ 19.45
$\pm$ 0.05 mag and turn off colours (B$-$V)$_{TO} =$ 0.39 $\pm$ 0.01 mag, (V$-$R)$_{TO}
=$ 0.27 $\pm$ 0.01 mag, (V$-$I)$_{TO} =$ 0.58 $\pm$ 0.01 mag. The turn off level is
slightly brighter than the value of V$_{TO} =$ 19.6 mag given by Alcaino et al. (1994)
and V$_{TO} =$ 19.7 mag given by Kravtsov et al. (1997). But it is slightly fainter
than the value of V$_{TO} =$ 19.4 mag in Harris et al. (1983). \\
The present CMDs give the value for HB magnitude V$_{HB} =$ 16.00 $\pm$ 0.10 mag. It is
0.$^{m}$15 and 0.$^{m}$25  brighter than the values given by Ferraro et al. (1992) and
Kravtsov et al. (1997) respectively. Thus we find $\Delta V_{TO-HB} =$ 3.45 $\pm$ 0.11
mag, close to the average value for Milky Way globular clusters (Buonanno et al. 1989).
The blue HB tail in NGC 1904 is one of the longest tail detected in GGCs. CMDs show that 
there is a HB bump never detected in an intermediate metal poor GGC. \\
A well defined sub giant branch, red giant branch, asymptotic giant branch, horizontal
branch, blue straggler sequence and main sequence are clearly visible in the apparent
CMDs (figure 2). \\
The giant branch is well defined in the CMDs. The tip of branch is located at V = 13.28
mag with colour (B$-$V) = 1.48 mag, (V$-$R) = 0.77 mag, (V$-$I) = 1.50 mag which is 
about 2.7 mag above the HB level. \\
The asymptotic giant branch is clearly visible in CMDs and it can be easily separated
from red giant branch upto the horizontal branch magnitude level. A well defined sequence
of blue straggler is seen above the main sequence turn off in all colours in CMDs (Figure 2).

\subsubsection {The Red Giant Branch and Metal Abundance}

The metalicity of a cluster can be estimated from CMD using the parameters $\Delta$V
(Sandage and Wallerstein 1960), S (Hartwick 1968) and (B$-$V)$_{0,g}$ (Sandage and Smith
1966). To determine $\Delta$V, S and (B$-$V)$_{0,g}$ cluster's interstellar reddening and
magnitude of zero age horizontal branch (ZAHB), V$_{HB}$ should be known. The interstellar
reddening of this cluster is almost negligible with E(B$-$V) $\sim$ 0.01 mag. We have not
calculated independent interstellar reddening of this cluster. Based on its earlier estimates,
we adopted the reddening value as E(B$-$V) = 0.01 for the cluster.  

\noindent
{\bf Table 5.} Metalicity determination from RGB parameters: relations and derived values
for NGC 1904.

\begin{center}
\begin{tabular}{ccl}\\\hline
Relation & Derived Metalicity & Reference \\
         &                     & (B$-$V)$_{0,g}$ \\\hline
4.30(B$-$V)$_{0,g} -$ 5.00 & $-$1.58 & Zinn and West 1984 \\
3.84(B$-$V)$_{0,g} -$ 4.63 & $-$1.57 & Gratton 1987 \\
4.68(B$-$V)$_{0,g} -$ 5.19 & $-$1.46 & Costar and Smith 1988 \\
2.85(B$-$V)$_{0,g} -$ 3.76 & $-$1.49 & Gratton and Ortolani 1989 \\ \hline
         &                     & $\Delta V_{1.4}$ \\\hline
$-$0.924 $\Delta V_{1.4} +$0.913 & $-$1.44 & Zinn and West 1984 \\
$-$1.01 $\Delta V_{1.4} +$1.30 & $-$1.28 & Costar and Smith 1988 \\
$-$0.65 $\Delta V_{1.4} +$0.28 & $-$1.38 & Gratton and Ortolani 1989 \\ \hline
         &                     &  S \\\hline
$-$0.29 S $-$0.01 & $-$1.78 & Gratton and Ortolani 1989 \\ \hline
\end{tabular}
\end{center}

\vspace*{0.3in}
\noindent
Adopting E(B$-$V) = 0.01, V$_{HB} =$ 16.00 $\pm$ 0.10 and the data in Table 5, we derived 
de reddened
colour of RGB at HB level (B$-$V)$_{0,g} =$ 0.80 mag; difference between the magnitude of the
RGB at a de reddened colour of (B$-$V)$_{0} =$ 1.4 mag and the HB $\Delta V_{1.4} =$ 2.55 mag and
slope of the RGB S = 6.09. Using these parameters we derived the values of [Fe/H] (table 5)
using different relations by different authors. The mean value of [Fe/H] obtained in this way are:\\
$[Fe/H]_{{(B-V)}_{0,g}} = -$1.52, $[Fe/H]_{{\Delta}_{V_{1.4}}} = -$1.37 and $[Fe/H]_{S} = -$1.78.
In conclusion, our metalicity estimation covers a range of $-$1.78 to $-$1.37 while previous
metalicity estimates shows (see Table 2 of Alcaino et al. 1994) the metalicity range for
this cluster is $-$1.78 to $-$1.42. Thus we confirm the previous metalicity determinations.

\subsubsection {Distance to the NGC 1904}

HB of a globular cluster is a standard candle for determine the distance modulus. We used
the relation between absolute magnitude of HB and metalicity given in the catalogue of
Harris (1994) for determining the M$_{V}$(HB) for the cluster
\begin{eqnarray}
M(HB) = 0.2 [Fe/H] + 1.0
\end{eqnarray}
We have used three different methods ((B$-$V)$_{0,g}, \Delta V_{1.4}$ and S) for calculating
the metalicity of the cluster. The mean value of metalicity determined from the above three
methods are $-$1.52, $-$1.37 and $-$1.78 respectively. Using the above equation we get the
following values of M(HB) for different values of metalicity:
\begin{eqnarray}
{[Fe/H] = -1.52, M(HB) = 0.70, (m-M) = 15.30} \\
{[Fe/H] = -1.37, M(HB) = 0.73, (m-M) = 15.27} \\
{[Fe/H] = -1.78, M(HB) = 0.64, (m-M) = 15.36}
\end{eqnarray}
The ishocrone fitting gives the value of (m$-$M) = 15.43 (see in section Ages of GGCs). Our
value for (m-M) is slightly less than the value reported by Ferraro et al. (1992) which is
15.45. They shifted the values of their fiducial sequences and overlap with M13. Alcaino et al.
(1994) have derived the value of (m-M) = 15.75 $\pm$ 0.1 mag while Stetson and Harris (1977)
have reported the value of (m-M) as 15.60. Both values are slightly larger than the value
obtained here.

\subsubsection {Asymptotic Giant Branch}

The AGB is the least studied of all sequences in GGCs CMDs. Generally this branch is not
clearly separated in most CMDs of GGCs. Reliable separation between RGB and AGB require
good accuracy of data points in CMDs. In the present study AGB is clearly separated from
RGB in all CMDs (figure 2.) for V in range from 13.2 to 15.6 mag.

\subsubsection {Blue Stragglers}

Blue Straggler Stars (BSSs) are found in all populations: in the field, in open clusters
of all ages (Population I, young disk, old disk), in globular clusters (Population II,
halo) and dwarf galaxies. These stars lie above the main sequence turn off in CMDs, a region
where, if BSSs had been normal single stars, they should already have evolved away from the
main sequence. These enigmatic stars lie blueward of turn off and appear to linger or 
straggler in their evolutionary process, hence the name blue stragglers. The actual positions
of BSSs in CMD may have a dependency on metalicity. (Fusi Pecci et al. 1992). 

\noindent
{\bf Table 6.} Fiducial sequences for V vs (B$-$V), (V$-$R) and (V$-$I) colour magnitude
diagrams of NGC 1904 displayed in Figure 2. Suffix 1, 2, 3 and 4 represent MS, SGB \& RGB,
AGB and HB respectively. 

\scriptsize
\noindent
\begin{tabular}{cccccccccccc} \\\hline
V$_{1}$ & (B$-$V)$_{1}$ & V$_{2}$ & (B$-$V)$_{2}$ & V$_{1}$ & (V$-$R)$_{1}$ & V$_{2}$ & (V$-$R)$_{2}$
& V$_{1}$ & (V$-$I)$_{1}$ & V$_{2}$ & (V$-$I)$_{2}$ \\\hline
19.50 & 0.40 & 18.22 & 0.64 & 19.46 & 0.28 & 18.89 & 0.35 & 19.49 & 0.58 & 18.80 & 0.76 \\
19.70 & 0.40 & 18.46 & 0.62 & 19.65 & 0.28 & 18.96 & 0.32 & 19.69 & 0.58 & 18.88 & 0.72 \\
19.86 & 0.41 & 18.68 & 0.60 & 19.80 & 0.28 & 19.06 & 0.30 & 19.84 & 0.59 & 18.93 & 0.68 \\
20.11 & 0.43 & 18.77 & 0.58 & 20.10 & 0.30 & 19.15 & 0.29 & 20.04 & 0.61 & 18.98 & 0.65 \\
20.29 & 0.45 & 18.86 & 0.55 & 20.48 & 0.31 & 19.27 & 0.28 & 20.25 & 0.63 & 19.08 & 0.61 \\
20.54 & 0.47 & 18.94 & 0.50 & 20.61 & 0.32 & 19.38 & 0.28 & 20.50 & 0.65 & 19.18 & 0.60 \\\cline{7-8}
20.73 & 0.50 & 19.01 & 0.45 & 20.84 & 0.33 & V$_{3}$ & (V$-$R)$_{3}$ & 20.70 & 0.67 & 19.31 & 0.58 \\\cline{7-8}\cline{11-12}
20.93 & 0.53 & 19.18 & 0.41 & 21.03 & 0.35 & 13.25 & 0.75 & 20.83 & 0.68 & V$_{3}$ & (V$-$I)$_{3}$ \\\cline{5-6}\cline{11-12}
21.10 & 0.56 & 19.34 & 0.40 & V$_{2}$ & (V$-$R)$_{2}$ & 13.46 & 0.72 & 21.00 & 0.71 & 13.26 & 1.43 \\\cline{1-6}\cline{9-10}
V$_{2}$ & (B$-$V)$_{2}$ & V$_{3}$ & (B$-$V)$_{3}$ & 13.33 & 0.76 & 13.67 & 0.68 & V$_{2}$ & (V$-$I)$_{2}$ & 13.46 & 1.37 \\\cline{1-4}\cline{9-10}
13.28 & 1.46 & 13.25 & 1.45 & 13.46 & 0.73 & 13.90 & 0.64 & 13.24 & 1.50 & 13.69 & 1.31 \\
13.45 & 1.38 & 13.41 & 1.37 & 13.69 & 0.69 & 14.05 & 0.62 & 13.36 & 1.45 & 13.97 & 1.23 \\
13.62 & 1.32 & 13.55 & 1.30 & 13.86 & 0.67 & 14.32 & 0.59 & 13.52 & 1.37 & 14.20 & 1.17 \\
13.74 & 1.27 & 13.74 & 1.24 & 14.07 & 0.64 & 14.66 & 0.54 & 13.72 & 1.32 & 14.41 & 1.12 \\
13.95 & 1.20 & 13.97 & 1.16 & 14.32 & 0.61 & 14.85 & 0.52 & 13.97 & 1.26 & 14.71 & 1.08 \\
14.16 & 1.14 & 14.17 & 1.08 & 14.66 & 0.57 & 15.11 & 0.49 & 14.25 & 1.20 & 14.89 & 1.03 \\
14.35 & 1.09 & 14.48 & 1.00 & 14.96 & 0.54 & 15.36 & 0.47 & 14.53 & 1.15 & 15.12 & 0.98 \\
14.54 & 1.05 & 14.72 & 0.87 & 15.21 & 0.53 & 15.61 & 0.44 & 14.74 & 1.11 & 15.37 & 0.94 \\\cline{11-12}
14.75 & 1.00 & 15.37 & 0.80 & 15.45 & 0.51 & 15.70 & 0.42 & 15.09 & 1.07 & V$_{4}$ & (V$-$I)$_{4}$ \\\cline{7-8}\cline{11-12}
14.97 & 0.96 & 15.61 & 0.70 & 15.68 & 0.50 & V$_{4}$ & (V$-$R)$_{4}$ & 15.32 & 1.04 & 15.98 & 0.35 \\\cline{3-4}\cline{7-8}
15.18 & 0.92 & V$_{4}$ & (B$-$V)$_{4}$ & 15.85 & 0.49 & 16.04 & 0.11 & 15.60 & 1.00 & 16.06 & 0.29 \\\cline{3-4}
15.35 & 0.89 & 16.12 & 0.14 & 16.04 & 0.48 & 16.18 & 0.08 & 15.85 & 0.98 & 16.18 & 0.22 \\
15.58 & 0.85 & 16.26 & 0.11 & 16.27 & 0.46 & 16.36 & 0.05 & 16.08 & 0.96 & 16.31 & 0.16 \\
15.94 & 0.81 & 16.50 & 0.06 & 16.50 & 0.45 & 16.61 & 0.01 & 16.34 & 0.93 & 16.49 & 0.11 \\
16.16 & 0.78 & 16.79 & $-$0.04 & 16.74 & 0.44 & 16.88 & $-$0.02 & 16.59 & 0.91 & 16.74 & 0.06 \\
16.35 & 0.76 & 17.12 & $-$0.12 & 16.97 & 0.43 & 17.03 & $-$0.03 & 16.87 & 0.89 & 17.07 & 0.01 \\
16.61 & 0.74 & 17.48 & $-$0.16 & 17.22 & 0.42 & 17.35 & $-$0.05 & 17.15 & 0.87 & 17.35 & $-$0.01 \\
16.87 & 0.72 & 17.88 & $-$0.18 & 17.50 & 0.42 & 17.67 & $-$0.06 & 17.38 & 0.86 & 17.71 & $-$0.05 \\
17.03 & 0.71 & 19.06 & $-$0.19 & 17.73 & 0.41 & 18.09 & $-$0.06 & 17.58 & 0.85 & 18.09 & $-$0.08 \\
17.27 & 0.69 &       &         & 17.98 & 0.40 & 18.26 & $-$0.06 & 17.86 & 0.84 &       &  \\
17.56 & 0.68 &       &         & 18.24 & 0.40 & 18.58 & $-$0.07 & 18.19 & 0.82 &       & \\
17.79 & 0.66 &       &         & 18.51 & 0.39 & 18.79 & $-$0.07 & 18.39 & 0.81 &       & \\
17.99 & 0.65 &       &         & 18.72 & 0.37 & 19.08 & $-$0.07 & 18.55 & 0.79 &       & \\\hline
\end{tabular}
\normalsize

\noindent
In the present study we have detected a sample of about 30 BSSs in magnitude range 17 $<$ V
$<$ 18.5 and colour range $-$0.2 $<$ (B$-$V) $<$ 0.5 (in Figure 1). Few BSSs of the present
sample are not seen in V vs (V$-$R) and V vs (V$-$I) CMDs due to our error rejection criterion.
Since BSSs are more centrally concentrated, the error $\delta$(V$-$R), $\delta$(V$-$I) is
more than 0.07 for BSSs not detected in these CMDs. 

\subsection {NGC 6341}

\subsubsection {Overall morphology of NGC 6341}

The V, (B$-$V) and V, (V$-$I) diagrams for all 3570 stars observed by us are presented in
Figure 3. The CMDs have well defined turn off (TO) region at V$_{TO} =$ 18.6 $\pm$ 0.05,
(B$-$V)$_{TO} =$ 0.42, (V$-$I)$_{TO} =$ 0.60. The values of HB parameters are V$_{HB} =$ 15.17
$\pm$ 0.05, (B$-$V)$_{HB} =$ 0.17, (V$-$I)$_{HB} =$ 0.30. Fiducial sequences of V vs (B$-$V)
and V vs (V$-$I) CMDs for MS, SGB and RGB are reported in Table 7 and plotted in figure 4 (a)
and figure 4 (b) respectively. V vs (B$-$V) and V vs (V$-$I) fiducial points published in 
the previous studies
of the same cluster are also plotted in figure 4 for comparison. \\
The CMDs of NGC 6341 reveal a relatively steep RGB and a HB population of the blue side of
RR Lyrae instability strip. Both of these features are characteristic of a metal poor stellar
system. \\
Previous studies (see table 3) show that the E(B$-$V) values for the cluster range from 0.02
to 0.03 mag. So, we adopt the value of E(B$-$V) = 0.02 mag for the cluster in the present
analysis. Keeping the same distance modulus value for V vs (V$-$I) CMD which we reported
for V vs (B$-$V) CMD, the best fit of fiducial sequence to theoretical isochrones give the
value of E(V$-$I) = 0.05 $\pm$ 0.01. \\
As noted above the HB of NGC 6341 is predominantly blueward of RR Lyrae instability strip.
The value of V$_{HB} =$ 15.17 $\pm$ 0.05. Figure 3 gives the following values of the RGB
colour at the level of HB: (B$-$V)$_g =$ 0.72. With the adopted reddening of E(B$-$V) = 0.02
this gives the value of (B$-$V)$_{0,g} =$ 0.70. Using the relation given by Gratton (1989)
{[Fe/H] = (2.85 $\pm$ 0.37) (B$-$V)$_{0,g} -$(3.76 $\pm$ 0.31)} which has rms error of 0.17 dex,
we get {[Fe/H] $= -$1.77 $\pm$ 0.20. Previous metalicity determinations of this cluster range
from $-$1.77 to $-$2.24 (Buonanno et al. 1985).
 
\noindent
{\bf Table 7.} Fiducial sequences for V vs (B$-$V) and (V$-$I) colour magnitude diagrams
of NGC 6341 displayed in Figure 3. Suffix 1, 2 and 3 represent MS, SGB \& RGB and HB respectively.

\scriptsize
\begin{center}
\begin{tabular}{cccccccc}\\\hline
V$_{1}$ & (B$-$V)$_{1}$ & V$_{2}$ & (B$-$V)$_{2}$ & V$_{1}$ & (V$-$I)$_{1}$ & V$_{2}$ & (V$-$I)$_{2}$
\\\hline
18.61 & 0.42 & 16.11 & 0.73 & 18.64 & 0.60 & 18.17 & 0.65 \\
18.71 & 0.42 & 16.46 & 0.70 & 18.89 & 0.61 & 18.44 & 0.61 \\
18.97 & 0.44 & 16.70 & 0.69 & 19.02 & 0.62 & 18.46 & 0.60 \\
19.19 & 0.45 & 16.93 & 0.68 & 19.29 & 0.64 & 18.57 & 0.60 \\\cline{7-8}
19.40 & 0.47 & 17.21 & 0.66 & 19.36 & 0.65 & V$_{3}$ & (V$-$I)$_{3}$ \\\cline{7-8}
19.57 & 0.49 & 17.49 & 0.65 & 19.60 & 0.67 & 15.18 & 0.30 \\
19.75 & 0.51 & 17.71 & 0.63 & 19.78 & 0.68 & 15.26 & 0.22 \\
19.92 & 0.53 & 17.90 & 0.61 & 20.05 & 0.72 & 15.45 & 0.14 \\
20.11 & 0.56 & 18.01 & 0.57 & 20.32 & 0.76 & 15.70 & 0.06 \\
20.26 & 0.59 & 18.11 & 0.51 & 20.58 & 0.80 & 16.00 & $-$0.01 \\
20.37 & 0.61 & 18.20 & 0.48 & 20.85 & 0.86 & 16.29 & $-$0.06 \\
20.56 & 0.65 & 18.35 & 0.44 & 21.00 & 0.89 & 16.50 & $-$0.11 \\
20.73 & 0.68 & 18.50 & 0.43 & 21.25 & 0.94 &   & \\\cline{3-4}
20.88 & 0.71 & V$_{3}$ & (B$-$V)$_{3}$ & 21.43 & 0.98 &   & \\\cline{3-4}
21.06 & 0.74 & 15.17 & 0.17 & 21.52 & 1.00 &    &   \\
21.21 & 0.78 & 15.26 & 0.11 & 21.74 & 1.06 &    &   \\
21.38 & 0.81 & 15.38 & 0.07 & 21.88 & 1.09 &    &   \\
21.62 & 0.86 & 15.57 & 0.01 & 22.12 & 1.15 &    &   \\\cline{5-6}
21.81 & 0.90 & 15.85 & $-$0.02 & V$_{2}$ & (V$-$I)$_{2}$ &   & \\\cline{5-6}
22.00 & 0.93 & 16.08 & $-$0.06 & 14.20 & 1.09 &    & \\
22.22 & 0.97 & 16.42 & $-$0.09 & 14.42 & 1.04 &    & \\\cline{1-2}
V$_{2}$ & (B$-$V)$_{2}$ &     &     & 14.76 & 1.01 &   &   \\\cline{1-2}
14.05 & 0.97 &      &     & 14.87 & 0.99 &    &    \\
14.20 & 0.94 &      &     & 15.16 & 0.96 &      &     \\
14.31 & 0.92 &      &     & 15.43 & 0.94 &      &     \\
14.48 & 0.90 &      &     & 15.74 & 0.91 &      &     \\
14.66 & 0.87 &      &     & 16.19 & 0.88 &      &     \\
14.87 & 0.85 &      &     & 16.41 & 0.87 &      &     \\
15.02 & 0.83 &      &     & 16.79 & 0.85 &      &     \\
15.24 & 0.81 &      &     & 17.06 & 0.84 &      &     \\
15.45 & 0.78 &      &     & 17.37 & 0.82 &      &     \\
15.66 & 0.76 &      &     & 17.75 & 0.79 & &      \\
15.88 & 0.74 &      &     & 18.04 & 0.72 &      &     \\\hline
\end{tabular}
\end{center}

\normalsize
\subsubsection {Distance Modulus}

Globular cluster distance can be determined by comparing the magnitude of stars on the
MS to a sample of subdwarfs with distance determine via parallax. This technique requires
fairly deep photometry to reach sufficiently far down the MS of a given GGC. Since the MS
is very well defined in a range of 3 to 3.5 mag fainter than MS turn off, we used best
fit of fiducial points to local subdwarfs to determine the distance modulus. We choose nearby
subdwarfs with good trigonometric parallaxes. \\
We used the five subdwarfs with best estimates of absolute magnitudes and colours. Parameters
for these stars are reported in Stetson and Harris (1988) and Laird et al. (1988). Data for
these stars are given in the Table 8. \\
The colour of each star was corrected for adopted reddening, leaving the distance modulus (m-M)$_{V}$
as free parameter. The fiducial sequence of NGC 6341 was then shifted until a satisfactory fit
was obtained. From visual matching of the two sequences, we derived (m-M)$_{V} =$ 14.48 $\pm$ 0.05.
 
\noindent
{\bf Table 8.} Selected field subdwarfs from those reported in Stetson and Harris (1988) (a)
and Laird et al. (1988) (b).

\begin{tabular}{rcccccc}\\\hline
HD & M$_{{V}_{a}}$ & (B$-$V)$_{a}$ & {[Fe/H]$_{a}$} & M$_{{V}_{b}}$ & (B$-$V)$_{b}$ & {[Fe/H]$_{b}$} \\\hline
25329 & 7.16 & 0.86 & $-$1.33 & 7.17$\pm$0.20 & 0.865 & $-$1.33 \\
103095 & 6.77 & 0.75 & $-$1.36 & 6.76$\pm$0.09 & 0.750 & $-$1.36 \\
134439 & 7.04 & 0.76 & $-$1.40 & 6.98$\pm$0.18 & 0.760 & $-$1.46 \\
134440 & 7.41 & 0.87 & $-$1.52 & 7.36$\pm$0.18 & 0.850 & $-$1.54 \\
201891 & 5.10 & 0.51 & $-$1.42 & 5.43$\pm$0.32 & 0.510 & $-$1.42 \\\hline
\end{tabular}

\section {ADD STAR EXPERIMENT: A NEW APPROACH}

In any discussion regarding the distribution of stars in crowded field
photometric regions of the sky as a function of magnitude, it is necessary
to first understand the effects of data completeness as a function of magnitude.
In general, it is easy to observe the brighter stars compare to the fainter stars.
If the observed stellar image is of a crowded field of the sky then it is necessary
to determine how the degree of completeness varies with crowding. Crowded regions
of the sky have higher density of stars and several stars are contaminated by
other stars. \\
To determine the true luminosity function from an uncorrected luminosity
function it is necessary to perform artificial stars tests to determine
the completeness correction and remove the field stars contamination. For
GGC NGC 6341 data we have done artificial stars tests and field stars
removal and written below in detail in this section of the paper. For
artificial stars tests we have suggested a new method and perform the
test on GGC NGC 6341 data in the present paper. \\
How to determine more accurate completeness factor? It is still a question of
debate. For this, different authors have suggested different methodologies.
A very simple method for this experiment is suggested by Mateo \& Hodge
(1987) and Mateo (1988). 
A brief description of the method is described here. Randomly selected artificial 
stars with a known magnitude range and random or known positions in the original 
image frames were added. Artifical stars can be generated in the observed image 
frames by using empirically derived point spread function (PSF) from each image 
frame using ADDSTAR routine in DAOPHOT II software. Then the image frames with 
artificially added stars are subjected to identical data reduction process 
using same PSF as the original frames were done. \\
Data completeness factor CF is defined by
\begin{eqnarray}
CF = {\frac {N_{recovered}} {N_{added}}}
\end{eqnarray}
Authors who have used the method suggested by Mateo \& Hodge (1987) and
Mateo (1988) believe that the procedure of the added artificial stars should
not change the crowding characteristics of the
original data frame by limited number ($\sim$ 5$-$15 \% of the number of
originally detected stars) can be added at one time. Thus, to have a satisfactory
number statistics for the determination of CF, one must has to repeat the above
process on a given data frame for many time. It requires a huge amount of the
CPU time. \\
We believe if the observed open cluster is very crowded or if it is a
galactic globular cluster then adding $\sim$ 5$-$15 \% stars of the originally
detected stars will change the crowding effect and hence, the PSF of the
original image frame and the image frame with added stars will vary
significantly if we redetermine the PSF of artificially added stars image
frame. Since artificial stars were generated by using original image PSF,
people use the PSF of original image frames on stars added image frame for
recovering the added artificial stars. \\
Since our final photometry file of the cluster take data from all images
frames (short, medium and long exposures at different air masses). We
generated artificial stars on all image frames of all passbands in the
following way: \\
{\bf 1.} For generating artificial stars we used ADDSTAR routine of
DAOPHOT II software. \\
{\bf 2.} For individual image frame we check stars detection magnitude
limit and then decide the magnitudes of artificial stars.  \\
{\bf 3.} We select the magnitude bin for determining the completeness
factor (as example 0.25 magnitude bin). In a box of 40 pixel $\times$
40 pixel we generate one star on every image frame. On one image frame
of CCD chip size of 320 pixel $\times$ 512 pixel, total 96 artificial
stars are generated. This type of five frames are generated for one
magnitude bin at one original image frame. The magnitude of
these stars on the five frames will vary by 0.05 magnitude which can
be used as the central magnitude of magnitude bin limit.
Using this method we can give equal weight to all artificially added
stars which will show real crowding effect. 40 pixel $\times$ 40 pixel
box will not overlap with two neighbor artificial stars. 
In the final photometry file of
NGC 6341 we have 3570 stars detected. So, in one image frame 96
artificial stars will not affect its crowding. \\
{\bf 4.} After generating artificial stars of different magnitude
on a specific image frame. We do the photometry of these new frames
using the PSF of the original image frame in the same manner as the
original image frames. This process will be applied to all original
image frames. \\
{\bf 5.} After doing the photometry on all new frames we determine
the ratio of total number of recovered stars and artificially generated
stars from all image frames for a specific magnitude bin and a specific
passband. It gives the completeness factor for the central magnitude of
the specific magnitude bin and specific passband. This process was
applied for all passbands, all magnitude bins and all image frames. \\
The completeness factor determined in this way for the main sequence stars
of the colour magnitude diagram of the GGC NGC 6341 is given in Table 9.

\noindent
{\bf Table 9.} Completeness factor for the cluster NGC 6341 are presented.
CF(B), CF(V) and CF(I) denote the completeness factor on image frames B, V
and I respectively. Magnitude denote the calibrated magnitude for B, V and
I pass bands.
\begin{center}
\begin{tabular}{cccc} \\\hline
{\bf Magnitude} & {\bf CF(B)}  & {\bf CF(V)} & {\bf CF(I)} \\\hline
18.00           & 0.96         & 0.91        & 0.57   \\
18.25           & 0.94         & 0.91        & 0.56   \\
18.50           & 0.92         & 0.89        & 0.52    \\
18.75           & 0.92         & 0.87        & 0.55    \\
19.00           & 0.89         & 0.82        & 0.52    \\
19.25           & 0.85         & 0.75        & 0.45    \\
19.50           & 0.79         & 0.64        & 0.41     \\
19.75           & 0.75         & 0.42        & 0.40     \\
20.00           & 0.65         & 0.26        & 0.35    \\
20.25           & 0.55         & 0.18        & 0.23     \\
20.50           & 0.40         &             & 0.16    \\
20.75           & 0.26         &             &          \\
21.00           & 0.14         &             &           \\\hline
\end{tabular}
\end{center}

\section {LUMINOSITY FUNCTION OF NGC 6341}

The luminosity function of a GGC can be used to study the cluster's state of
dynamical evolution and initial mass function; together these two factor lead
to the cluster's present day mass function. Since in our data we don't have
sufficient mass range in masses of main sequence stars, so, we were not able
to determine the present day mass function of the cluster.
From the main sequence of the colour magnitude diagram, luminosity function
of NGC 6341 were determined from star counts in the bin width of 0.25 mag in V passband.
The first step towards the estimation of main sequence luminosity function
is the estimation of the field star contamination. The contamination by the
field stars in the direction of NGC 6341 has been estimated by Ratnatunga
\& Bahcall (1985) using the Galaxy model given by Bahcall \& Soneria (1980).
The expected number of field stars in different magnitude bins with different
colours are given in table 10.  

\noindent
{\bf Table 10.} Field star contamination correction in the observed galactic
globular cluster NGC 6341 field according to model of Ratnatunga \& Bahcall
(1985).
\begin{center}
\begin{tabular}{ccccc}\\\hline
           & V $<$ 15 & 15 $<$ V $<$ 17 & 17 $<$ V $<$ 19 & 19 $<$ V $<$ 21 \\\hline
B $-$ V $<$ 0.8 & 1 & 2 & 4 & 7 \\
0.8 $<$ B $-$ V $<$ 1.3 & 0 & 2 & 4 & 3 \\
B $-$ V $>$ 1.3 & 0 & 0 & 3 & 10 \\\hline
\end{tabular}
\end{center}

\noindent
The main factor which limits the precise determination of luminosity function
from observation is data completeness. Data completeness can be estimated by
artificial add star experiment which is done in the previous section of the paper
by our new method. \\
The main sequence luminosity function is derived for a bin width of 0.25 mag
in V using stars in V vs (B$-$V) colour magnitude diagram. The procedure used
for the data completeness correction has been described in detail earlier by
Sagar \& Richtler (1991) and again recently by Sagar \& Griffiths (1998b).
Luminosity function of NGC 6341 is given in the table 11.  \\
Figure 3 shows the luminosity function derived from the original star counts
and after the completeness correction and field star subtraction. The slope
of the luminosity function is derived as 0.58$\pm$0.05 from the histogram
plotted in the figure 6. 

\newpage
\noindent
{\bf Table 11.} Luminosity function of NGC 6341. The counts are taken from
the V Vs (B $-$ V) colour magnitude diagram while the completeness factor is
taken as minimum of the pair (CF(B), CF(V)). RN denotes star counts in 0.25
magnitude bin width. After the correction of data completeness and field star
contamination RN yields N.

\begin{center}
\begin{tabular}{ccrcrc}\\\hline
V(mag) & CF & RN & log(RN) & N  & log(N) \\\hline
18.00  & 0.91 & 21 & 1.32 & 22.08 & 1.34 \\
18.25  & 0.91 & 32 & 1.51 & 34.16 & 1.53 \\
18.50  & 0.89 & 41 & 1.61 & 45.07 & 1.65 \\
18.75  & 0.87 & 66 & 1.82 & 74.36 & 1.87 \\
19.00  & 0.82 & 59 & 1.77 & 70.95 & 1.85 \\
19.25  & 0.75 & 76 & 1.88 & 99.33 & 2.00 \\
19.50  & 0.64 & 94 & 1.97 & 144.87 & 2.16 \\
19.75  & 0.42 & 131 & 2.12 & 309.90 & 2.49 \\\hline
\end{tabular}
\end{center}

\section {AGES OF THE GALACTIC GLOBULAR CLUSTERS}

The ages of globular clusters has been a topic of great interest for many years. The primary
reason for this interest is cosmological: the universe must be older than the objects within
it. Milky Way globular clusters contains the oldest stars for which reliable age estimates
are available, and thus provide a lower limit of the age of the Universe. Ages of GGCs are
debated since long because the lower limit of the age of the universe on the basis of GGCs
ages more than the upper limit of the age of the Universe on the basis of some cosmological
models. \\
The well studied GGCs NGC 1904 and NGC 6341 are excellent objects for
age determinations. The have high galactic latitudes, so very little interstellar redenning,
and low metalicities. The latter is helpful since the stellar models are more reliable at
low metalicity. \\
In order to determine the age of a cluster by theoretical isochrone fitting, one should know
about the main sequence turn off (MSTO) point, distance modulus, redenning of the cluster.
These parameters are derived for these GGCs in the previous sections.

\subsection {NGC 1904}

The determination of the distance modulus by fitting the isochrones to fiducial sequences of
CMDs of the cluster NGC 1904 gives the value of (m-M)$_{V} =$ 15.43 mag for redenning values of
E(B$-$V) = 0.01 mag, E(V$-$R) = 0.02 mag, E(V$-$I) = 0.05 mag. \\
In order to derive age of the cluster stars, we converted apparent V, (B$-$V), (V$-$R) and
(V$-$I) diagrams into intrinsic ones. Figure 7 displays plots of the absolute magnitude M$_{V}$
against (B$-$V)$_{0}$, (V$-$R)$_{0}$ and (V$-$I)$_{0}$. We have estimated the age of the cluster
stars by fitting stellar evolutionary isochrones appropriate for the cluster metalicity given
by Demarque et al. (1996) to the fiducial sequences of NGC 1904, we obtained the
age of the cluster as 14$\pm$1 GYr. \\
Previous age estimation of this cluster has a ranges from 12 to 19 Gyr (table 2). Our age
estimation is about the average of previous age determination of this GGC.

\subsection {NGC 6341}

In order to derive age of the cluster stars, we converted apparent V, (B$--$V) and (V$-$I)
diagrams into intrinsic ones. Using the value of distance modulus (m-M)$_{V} =$ 14.48 and
redenning E(B$-$V) = 0.02 and E(V$-$I) = 0.05. Figure 8 displays plots of the absolute
magnitude M$_{V}$ against (B$-$V)$_{0}$ and (V$-$I)$_{0}$. We have estimated the age of the
cluster by fitting stellar evolutionary isochrones given by Demarque et al. (1996) for
the cluster metalicity in figure 8. By fitting the isochrones to the fiducial sequences
of NGC 6341, we obtained the age of the cluster is 16$\pm$1 GYr. 

\section {RELATION BETWEEN METALICITY AND STRUCTURAL PARAMETERS OF GGCs}

Using main sequence luminosity functions of 9 galactic globular clusters
(hereafter GGCs), McClure et al. (1986) suggested that the mass function
power law index x of GGCs depends on the metalicity [Fe/H] of the cluster.
Pryor et al. (1986) calculated mass segregation corrections for GGCs mass
functions using multicomponent King models with power law mass functions
to eight observed GGCs mass functions. They found that the corrected mass
function exponents still show the previously reported intrinsic variation
among GGCs and the correlation of that variation with metalicity. Deep 
luminosity function
studies of the GGC M30 is done by Piotto et al. (1990) using 2.2 meter ESO/MPI
telescope and CCD detector. Their data agree well with the trend of x found
by Pryor et al. (1986) and yielded a global mass function x$_{0} \sim$ 0.7. \\
Capaccioli et al. (1991) took the sample of 14 GGCs with deep CCD
luminosity functions and found that formally suggested dependence of x
on [Fe/H] is not supported by the new data. Capaccioli et al. (1991)
found a possible trend of the slope with the galactocentric distance R
and with the distance from the galactic disk Z$_g$ of GGCs.
Using original luminosity functions of 17 GGCs data, Capaccioli et al. (1993)
derived the mass functions of these GGCs in a consistent way. They interpret
the dependence of the mass function slope on the R and Z$_g$ as an evidence
of a selective loss of stars induced by the GGC dynamical evolution. The same
data of a sample of 17 GGCs by Capaccioli et al. (1993) is used by Djorgovski
et al. (1993). They analyzed the dependence of stellar mass function slopes for
a sample of 17 GGCs on different parameters of GGCs. They found the mass function
slopes in the range of 0.5 $\leq (M / {M_{\odot}}) \leq$ 0.8 are largely determined
by R and Z$_{g}$ and also a lesser extent on the GGCs metalicity. Other parameters
of GGCs have little effect on the mass function slopes. \\
Drukier et al. (1988) have done deep luminosity function studies of GGC M13
using CFHT and CCD detector. They found the shape of mass function of this
cluster is inconsistent with the predicted for mass segregation near the
position of M13 in the models of Pryor et al. (1986). Mass functions studies of 
2 GGCs M13 and M71 down to the main sequence to 0.2
M$_{\odot}$ was done by Richer et al. (1990). They used existing data of GGC
NGC 6397 also for covering entire range in metal abundance of GGCs. They reported
there is no correlation between the mass function slopes and the cluster metal
abundance. Based on observations with the NTT at ESO and the du Pont Telescope
at Las Campanas Observatory Richer et al. (1991) determined luminosity and
mass function of GGCs $\omega$ Cen, M5 and NGC 6752. They used their new data
of these GGCs and existing data for GGCs M13, NGC 6397 and M71 to investigate
any systematics of the cluster mass function slopes and their evolution. They
reported a very important and quite robust result is that some (and perhaps
all) clusters probably have very steep initial mass function slope exceeding
2.5 (Salpeter value 1.3). No clear dependence of the slope on the GGCs structural
parameters or metal abundance is found by these authors.  \\
In the present work we took the data of 64 GGCs from Djorgovski (1993).
These GGCs parameters are very accurately known. We plotted log of
galactocentric distance of GGCs (log R Vs [Fe/H]) and log of distance from
the galactic plane of GGCs (log Z$_{g}$ Vs [Fe/H]) and found there are clear
trend in these two plots. So, we decided to divide all GGCs in five metalicity
bins and each bin should contain $\sim$ 20\% of the GGCs from the sample.
We plotted mean and median of log R Vs [Fe/H] and log Z$_{g}$ Vs [Fe/H] in the
figure 9. Using least square fit we got the following relations: \\
For Mean Value of log R, log Z$_g$ and [Fe/H]
\begin{eqnarray}
log \ R = -0.344 \ [Fe/H] + 0.297 \\
log \ Z_{g} = -0.390 \ [Fe/H] - 0.144
\end{eqnarray}
For Median Value of log R, log Z$_g$ and [Fe/H]
\begin{eqnarray}
log \ R = -0.226 \ [Fe/H] + 0.462 \\
log \ Z_{g} = -0.353 \ [Fe/H] - 0.105
\end{eqnarray}
\noindent
{\bf Table 12.} Mean and Median of Globular Cluster Parameters \\
\begin{tabular}{ccccc} \\\hline
            &   Mean &              & Limits of   & No. of \\
  log(R)    &  log Z$_{g}$  &  [Fe/H]      & [Fe/H]      & GGCs   \\\hline
0.525$\pm$0.294 & 0.099$\pm$0.298 & $-$0.604$\pm$0.176 & \hspace*{0.6in} [Fe/H] $> -$1.00 & 13 \\
0.809$\pm$0.368 & 0.453$\pm$0.484 & $-$1.318$\pm$0.071 & $-$1.40 $\leq$ [Fe/H] $< -$1.00 & 12 \\
0.748$\pm$0.477 & 0.389$\pm$0.513 & $-$1.520$\pm$0.060 & $-$1.60 $\leq$ [Fe/H] $< -$1.40 & 15 \\
0.779$\pm$0.328 & 0.391$\pm$0.456 & $-$1.691$\pm$0.067 & $-$1.80 $\leq$ [Fe/H] $< -$1.60 & 13 \\
1.087$\pm$0.468 & 0.733$\pm$0.549 & $-$2.019$\pm$0.137 & \hspace*{0.6in} [Fe/H] $< -$1.80  & 11 \\\hline
            &   Median      &              & Limits of   & No. of \\
  log(R)    &  log Z$_{g}$  &  [Fe/H]      & [Fe/H]      & GGCs   \\\hline
0.600$\pm$0.294 & 0.110$\pm$0.298 & $-$0.590$\pm$0.176 & \hspace*{0.6in} [Fe/H] $> -$1.00 & 13 \\
0.880$\pm$0.368 & 0.525$\pm$0.484 & $-$1.315$\pm$0.071 & $-$1.40 $\leq$ [Fe/H] $< -$1.00 & 12 \\
0.710$\pm$0.477 & 0.260$\pm$0.513 & $-$1.540$\pm$0.060 & $-$1.60 $\leq$ [Fe/H] $< -$1.40 & 15 \\
0.710$\pm$0.328 & 0.370$\pm$0.456 & $-$1.670$\pm$0.067 & $-$1.80 $\leq$ [Fe/H] $< -$1.60 & 13 \\
1.020$\pm$0.468 & 0.730$\pm$0.549 & $-$2.019$\pm$0.137   & \hspace*{0.6in} [Fe/H] $< -$1.80  &  11  \\\hline
\end{tabular}

\section {CONCLUSIONS}

The main conclusions of the present paper are as follows: \\
{\bf 1.} Using B,V,R,I photometry of NGC 1904, we derived E(V$-$R) = 0.02 mag, E(V$-$I) = 0.05 mag,
V$_{HB}$ = 16.00$\pm$0.10 mag, (B$-$V)$_{g} =$ 0.81 mag, (B$-$V)$_{0,g} =$ 0.80 mag. Metalicity
[Fe/H] = $-$1.52, $-$1.37 and $-$1.78 were derived by using (B$-$V)$_{0,g}$, $\Delta V_{1.4}$
and S methods respectively. (m$-$M)$_{V}$ = 15.30, 15.27 and 15.36 were derived by
using (B$-$V)$_{0,g}$, $\Delta V_{1.4}$ and S methods respectively. The values of morphological
parameters of the cluster are V$_{TO} =$ 19.45$\pm$0.05 mag, (B$-$V)$_{TO} =$ 0.39$\pm$0.01 mag, 
(V$-$R)$_{TO} =$ 0.27$\pm$0.01 mag,
(V$-$I)$_{TO} =$ 0.58$\pm$0.01 mag and $\Delta V_{TO-HB} =$ 3.45$\pm$0.11 mag. \\
{\bf 2.} Using B,V,I photometry of NGC 6341, we derived E(V$-$I) = 0.05,
V$_{HB}$ = 15.17$\pm$0.05 mag, (B$-$V)$_{HB} =$ 0.17 mag, (V$-$I)$_{HB} =$ 0.30 mag, (B$-$V)$_{g} =$ 0.72 mag, 
(V$-$I)$_{g} =$ 0.86 mag, (B$-$V)$_{0,g} =$ 0.70 mag, (V$-$I)$_{0,g} =$ 0.81 mag, metalicity
[Fe/H] = $-$1.77$\pm$0.20 by using (B$-$V)$_{0,g}$, (m$-$M)$_{V}$ = 14.48$\pm$0.05 mag from 
subdwarf fitting. The value of V$_{TO} =$ 18.60$\pm$0.05 mag, (B$-$V)$_{TO} =$ 0.42 mag and 
(V$-$I)$_{TO} =$ 0.60 mag and $\Delta V_{TO-HB} =$ 3.43$\pm$0.10 mag. \\
{\bf 3.} We have suggested a new, add star experiment technique. While its implementation
is still in the preliminary stage. We have successfully used it to determine the
completeness factors in B, V and I passbands data of GGC NGC 6341. In the near
future we will perform the same test on other GGCs data. 
Using our new method of artificial stars, we added a total of 141120 artificial
stars in 1470 separate runs of ADDSTAR routine of DAOPHOT II. For each of the
1470 artificial stars test runs, 96 stars were added to all 49 image frames of
B, V and I passbands of the GGC NGC 6341. The above process took about a week
long time and 100 \% CPU time on a Sun Sparc 1 workstation. \\
{\bf 4.} We determined the main sequence luminosity function of NGC 6341 down
to V $\sim$ 21.0 (M$_{V} \sim$ 6.5). The slope of luminosity function is derived
as 0.58$\pm$0.05. \\ 
{\bf 5.} By using theoretical isochrone fitting we determined the ages of GGCs NGC 1904 and 
NGC 6341 to 14.0$\pm$1.0 Gyr and 16.0$\pm$1.0 Gyr respectively. \\
{\bf 6.} In the present paper we have tried to solve the controversy
about the dependence of mass function power law index of GGCs on metalicity or 
R and Z$_g$. Figure 9 and equations 10$-$13 shows that there is clear trend in the 
plots log R Vs. [Fe/H] and log Z$_g$ Vs. [Fe/H]. We conclude that R, Z$_g$ and
[Fe/H] are not three independent parameters of GGCs but R and Z$_g$ depends
on [Fe/H]. \\
Till date we have mass function study of a few $\sim$ 20 \% GGCs only.
As sample of mass function measurements of GGCs will increase, we will have
better opportunity to obtain more accurate estimation on the empirical correlation
between GGCs fundamental parameters and their mass functions slope. At present
we can not say confidently its dependent on any fundamental parameter of GGCs.

\Large
\noindent
{\bf ACKNOWLEDGMENTS} \\ 
\normalsize
I am grateful to Prof. Ram Sagar for providing me his photometric
data of NGC 6341 (M92) and valuable suggestions. This work was supported in 
part by DST Project No. SP/S2/011/93 (PRU) dated 02.01.1997 and Department 
of Atomic Energy, Government of India.

\newpage
\noindent
\Large
{\bf FIGURE CAPTIONS} \\
\normalsize
{\bf Figure 1.} Errors in the photometric observations of NGC 1904 and
NGC 6341 as a function of brightness. \\
{\bf Figure 2.} V vs (B $-$ V), V vs (V $-$ R) and V vs (V $-$ I) colour 
magnitude diagrams of the galactic globular cluster NGC 1904. \\
{\bf Figure 3.} V vs (B $-$ V) and V vs (V $-$ I) colour magnitude diagrams
of the galactic globular cluster NGC 6341. \\
{\bf Figure 4. (a)} Comparison between the fiducial points of V vs (B $-$ V)
CMD of NGC 6341 from the present work (solid line), Sandage (1970) (dots), 
Heasley \& Christian (1986) (short dashed), Stetson \& Harris (1988) (long dash).
{\bf (b)} Fiducial points of V vs (V $-$ I) CMD of the present work (solid line)
and by Johnson \& Bolte (1998) (dots). \\
{\bf Figure 5.} Fit of the fiducial sequence of NGC 6341 to the field subdwarfs.
Left: data from Stetson \& Harris (1988); Right: data from Laird et al. (1988). \\
{\bf Figure 6.} Luminosity function of main sequence stars in V Vs (B$-$V)
colour magnitude diagram of NGC 6341. Dotted line represents the log of actual
counts of stars in the colour magnitude diagram. Continuous line represents
the log of count after applying completeness correction and field star
contamination. \\
{\bf Figure 7.} Ishocrone fitting to the fiducial points of NGC 1904. Fit with the
isochrone by Demarque et al. (1996). A good fit is obtained for age 14 Gyr for the
CMDs of the present study. \\
{\bf Figure 8.} Ishocrone fitting to the fiducial points of NGC 6341. Fit with the
isochrone by Demarque et al. (1996). A good fit is obtained for age 16 yr for the
CMDs of the present study. \\
{\bf Figure 9.} Data of the physical parameters of 64 GGCs from Djorgovski (1993)
are plotted. Mean of data is plotted by filled circle and solid line. Median of the
data is plotted by open circle and dotted line.  

\newpage
\noindent
\Large
{\bf REFERENCES} 

\noindent
\normalsize
Alcaino, G., 1976, A\&AS, 26, 353 \\
Alcaino, G., Liller, W., Alvarado, F., 1987, AJ, 93, 464 \\
Alcaino, G., Liller, W., Alvarado, F., Wenderoth, E., 1994, AJ, 107, 230 \\
Arp, H. C., Baum, W. A., Sandage, A. R., 1952, AJ, 57, 4 \\
Arp, H. C., Baum, W. A., Sandage, A. R., 1953, AJ, 58, 4 \\
Bahcall, J. N., Soneria, R. M., 1980, ApJS, 44, 73 \\
Buonanno, R., Corsi, C. E., Fusi Pecci, F., 1985, A\&A, 145, 97 \\
Buonanno, R., Corsi, C. E., Fusi Pecci, F., 1989, A\&A, 216, 80 \\
Capaccioli, M., Ortolani, S., Piotto, G., 1991, A \& A, 244, 298 \\
Capaccioli, M., Piotto, G., Stiavelli, M., 1993, MNRAS, 261, 819 \\
Corodoni, J. P., Auriere, M., 1983, A\&AS, 54, 431 \\
Costar, D., Smith, H. A., 1988, AJ, 96, 1925 \\
Demarque, P., Chaboyer, B., Guenther, D., et al. 1996, private communication \\
Djorgovski, S. G. 1993, in Structure and Dynamics of Globular Clusters by
S. G. \\
\hspace*{0.3in} Djorgovski \& G. Meylan eds. ASP conf. series 50, p. 373 \\
Djorgovski, S., Piotto, G., Capaccioli, M., 1993, AJ, 105, 2148 \\
Drukier, G. A., Fahlman, G. G., Richer, H. B., 1988, AJ, 95, 1415 \\
Ferraro, F. R., Clementini, G., Fusi Pecci, F., et al. 1992, MNRAS, 256, 391 \\
Ferraro, F. R., Messino, M., Fusi Pecci, F., et al. 1999, AJ, 118, 1738 \\
Fusi Pecci, F., Ferraro, F. R., Corsi, C. E., et al. 1992, AJ, 104, 1831 \\ 
Goranskij, V. P., 1976, Astron. Tsirk., 902, 5 \\
Gratton, R. G., 1987, A\&A, 179, 181 \\
Gratton, R. G., Ortolani, S., 1986, A\&AS, 65, 63 \\
Gratton, R. G., Ortolani, S., 1989, A\&A, 211, 41 \\
Gupta, A. C., Subramaniam, A., Sagar, R., Griffiths, W. K., 2000, A \& AS,
145, 365 \\
Hanes, D. A., Whittaker, D. G., 1987, AJ, 94, 906 \\
Harris, W. E., Pudritz, R. E., 1994, ApJ, 429, 177 \\
Harris, W. E., Racine, R., 1979, ARAA, 17, 241 \\
Harris, W. E., Hesser, J. E., Atwood, B., 1983, PASP, 95, 951 \\
Hartwick, F. D. A., 1968, ApJ, 154, 475 \\
Heasley, J. N., Christian, C. A., 1986, ApJ, 307, 738 \\
Heasley, J. N., Janes, K. A., Christian, C. A., 1986, AJ, 91, 1108 \\
Johnson, J. A., Bolte, M., 1998, AJ, 115, 693 \\
Kravtsov, V., Ipatov, A., Samus, N., et al. 1997, A\&AS, 125, 1 \\
Laird, J. B., Carney, B. W., Latham, D. W., 1988, AJ, 95, 1843 \\
Landolt, A. U., 1983, AJ, 88, 439 \\
Mateo, M., 1988, ApJ, 331, 261 \\
Mateo, M., Hodge, P., 1987, ApJ, 320, 626 \\
McClure, R. D., VandenBerg, D. A., Smith, G. H., Fahlman, G. G., Richer, H. B., \\
\hspace*{0.2in} Hesser, J. E., Harris, W. E., Stetson, P. B., Bell, R. A., 1986, ApJL,
307, L49 \\
McLaughlin, D. E., 1994, PASP, 106, 47 \\
Piotto, G., King, I. R., Capaccioli, M., Ortolani, S., Djorgovski, S., 1990,
ApJ, 350, \\
\hspace*{0.2in} 662 \\
Pryor, C., Smith, G. H., McClure, R. D., 1986, AJ, 92, 1358 \\
Racine, R., Harris, W. E., 1992, AJ, 104, 1068 \\
Ratnatunga, K. U., Bahcall, J. N., 1985, ApJS, 59, 63 \\
Richer, H. B., Fahlman, G. G., Buonanno, R., Fusi Pecci, F., 1990, ApJ, 359, L11 \\
Richer, H. B., Fahlman, G. G., Buonanno, R., Fusi Pecci, F., Searle, L.,
Thompson, \\
\hspace*{0.2in} I. B., 1991, ApJ, 381, 147 \\
Sagar, R., Richtler, T., 1991, A\&A, 250, 324 \\
Sagar, R., Griffiths, W. K., 1991, MNRAS, 250, 683 \\
Sagar, R., Griffiths, W. K., 1998a, MNRAS, 299, 1 \\
Sagar, R., Griffiths, W. K., 1998b, MNRAS, 299, 777 \\
Sandage, A., 1970, ApJ, 162, 841 \\
Sandage, A., 1983, AJ, 88, 1159 \\
Sandage, A., Katem, B., 1983, AJ, 88, 1146 \\
Sandage, A. R., Smith, L. L., 1966, ApJ, 144, 886 \\
Sandage, A., Walker, M. F., 1966, ApJ, 143, 313 \\
Sandage, A. R., Wallerstein, G., 1960, ApJ, 131, 598 \\
Secker, J., 1992, AJ, 104, 1472 \\
Secker, J., Harris, W. E., 1993, AJ, 105, 1358 \\
Stetson, P. B., 1987, PASP, 99, 191 \\
Stetson, P. B., 1992, IAU col. 136 on stellar photometry $-$ current techniques and
future developments, eds. C. J. Butler and I. Elliot, p. 291 \\
Stetson, P. B., Harris, W. E., 1977, AJ, 82, 954 \\
Stetson, P. B., Harris, W. E., 1988, AJ, 96, 909 \\
van den Bergh, S., 1985, ApJ, 297, 361 \\
Zinn, R. J., West, M. J., 1984, ApJS, 55, 45 \\
\end{document}